\documentclass[12pt,draftclsnofoot,onecolumn]{IEEEtran}

\ifCLASSINFOpdf
\else
\fi
\usepackage{graphicx}
\usepackage{epstopdf}
\usepackage{multicol}
\usepackage{amsmath}
\usepackage{mathtools}
\usepackage{amssymb}
\usepackage{esint}
\usepackage{amsfonts}
\usepackage{multirow}
\usepackage{wrapfig}
\usepackage{cite}
\usepackage{subcaption}
\usepackage{lipsum}
\usepackage{float}    
\usepackage{stfloats} 
\usepackage{algorithm}
\usepackage[noend]{algpseudocode}
\usepackage{caption}
\usepackage[none]{hyphenat}
\usepackage[utf8]{inputenc}
\usepackage[T1]{fontenc}
\usepackage[english]{babel} 
\usepackage{amsmath,amsfonts,amsthm,bm} 
\usepackage{nomencl}
\DeclareMathOperator*{\argmax}{argmax}

\usepackage{xcolor}
 
\begin{document}

\title{Deep Reinforcement Learning Powered IRS-Assisted Downlink NOMA}

\author{Muhammad Shehab,~\IEEEmembership{Student Member,~IEEE,}
        Bekir S. Ciftler,~\IEEEmembership{Member,~IEEE,}
        Tamer Khattab,~\IEEEmembership{Senior Member,~IEEE,}
        Mohamed M. Abdallah,~\IEEEmembership{Senior Member,~IEEE,}
        Daniele Trinchero.
        
\thanks{M. Shehab and T. Khattab are with the Department of Electrical Engineering, Qatar University, Doha, Qatar.  B. Ciftler and M. Abdallah are with the Division of Information and Computing Technology, College of Science and Engineering, Hamad Bin Khalifa University, Doha, Qatar.  D. Trinchero is with Dipartimento di Elettronica, Politecnico di Torino, Torino, Italy.}} 


\maketitle

\IEEEpeerreviewmaketitle

\begin{abstract}

In this work, we examine an intelligent reflecting surface (IRS) assisted downlink non-orthogonal multiple access (NOMA) scenario with the aim of maximizing the sum rate of users. The optimization problem at the IRS is quite complicated, and non-convex, since it requires the tuning of the phase shift reflection matrix. Driven by the rising deployment of deep reinforcement learning (DRL) techniques that are capable of coping with solving non-convex optimization problems, we employ DRL to predict and optimally tune the IRS phase shift matrices. Simulation results reveal that IRS assisted NOMA based on our utilized DRL scheme achieves high sum rate compared to OMA based one, and as the transmit power increases, the capability of serving more users increases. Furthermore, results show that imperfect successive interference cancellation (SIC) has a deleterious impact on the data rate of users performing SIC. As the imperfection increases by ten times, the rate decreases by more than 10\%. 

\end{abstract}


\section{Introduction}
\IEEEPARstart{W}{ith}  the   increasing   progressions   in   the   wireless communications,   future   communication systems   are   favorable to support  higher  data  rates,  higher  spectral  efficiencies,  reduced latencies,   wider   coverage   areas   and   massive number of connections,  among  other  features.

\IEEEpeerreviewmaketitle


As  wireless  technologies have grown exponentially over the last few decades, wireless systems are promising to meet the demand for the enormous number of connections. The next generation networks will be an end-to-end ecosystem to enable a fully connected and sustainable community.  The main purpose of these networks is to provide seamless  and  ubiquitous communications for users  with a higher throughput, low latency, low energy consumption and support the escalation in mobile  data  consumption  for  hundreds  of  thousands  of connections. As 5G networks are being deployed, technologies for 6G networks are being researched and examined to attain more reliable and faster communication systems \cite{IEEEhowto:1}.

Among these technologies, are the intelligent reflecting surfaces (IRS)s, which regulate the wireless environment to boost the energy and spectral efficiencies. IRS consists of a huge number of passive elements or IRS units, each unit can passively reflect the incident electromagnetic wave signal and modify it in terms of phase, frequency, amplitude or polarization. Most of the research papers in the literature are considering passive IRS where only a phase shift to the incident signal is applied. Thus, the IRS will not consume any transmit power. Consequently, the IRS optimization problem is focused on the phase-shift matrix \cite{IEEEhowto:2}. Further, the IRS can aid the transmissions among the transmitters and receivers, especially where there is no line of sight (LoS) between the transmitting antenna and the receiving antenna, or if the direct link suffers from shadowing and deep fading rendering the quality of the channel for direct communications unreliable \cite{IEEEhowto:3}. Compared to decode and forward and amplify and forward, IRS demands less energy and power consumption because of its passive features. Hence, IRS is anticipated to be a promising solution for future 5G / 6G communication systems \cite{IEEEhowto:4}. IRS can be related to technologies such as massive or large multiple input multiple output (M-MIMO) antenna networks since it utilizes large number of antennas to increase the energy and spectrum efficiency. Thus, IRS is considered a potential component in 6G networks, which is analogous to M-MIMO potential in 5G networks. Nonetheless, the difference is that the IRS regulates the propagation in the wireless environment. 

Non-orthogonal multiple access (NOMA) is a vital component in 5G wireless communication systems and beyond because of the high spectrum efficiency it provides in addition to its support for massive connectivity. In the previous cellular systems, many multiple access technologies were adopted such as the time division multiple access (TDMA), frequency  division multiple access (FDMA), spatial division multiple access (SDMA) and orthogonal frequency division multiple access (OFDMA). Based on their design, these technologies are considered as orthogonal multiple access (OMA) techniques, since the wireless resources are allocated to multiple users orthogonally. The users are separated in the chosen access domain whether it is in frequency, time, or space. If orthogonality is violated, the users will suffer from interference and quality of communications links will degrade leading to loss of information and/or inefficient resources utilization. Nonetheless, OMA schemes cannot satisfy the requirements for future communication systems which causes the need for NOMA \cite{IEEEhowto:5}. NOMA  achieves high sum rate capacity as compared to the traditional orthogonal multiple access (OMA) techniques. The reason is that it enables multiple users to transmit simultaneously in the same set of shared resources. This results in an interference, but NOMA utilizes a method called successive interference cancellation (SIC) to eliminate the resulting interference.
 
\subsection{Related Work}

Many research studies related to IRS in 5G and 6G communication networks are being conducted. Some of these studies inspected IRS for OMA schemes such as in \cite{IEEEhowto:6} - \cite{IEEEhowto:10}, but these studies did not include NOMA in their scenarios. However, several recent research studies started investigating IRS for NOMA communication. The authors in \cite{IEEEhowto:11} examined an uplink scenario for IRS NOMA and they maximized the sum rate for all users taking into consideration the power constraint for each user. They solved the non-convex problem using semi-definite relaxation which is a mathematical method and provides near-optimal solution. In \cite{IEEEhowto:12}, the authors proposed a simple design for the IRS assisted NOMA system. First they utilized spatial division multiple access on the base station (BS) side for orthogonal beams generation. This is done by making use of the spatial  directions of the channels of the nearby users. Then the authors employed IRS aided NOMA to serve the additional users on the cell edge. They demonstrated the performance of the IRS-NOMA system by providing analytical results. The authors in \cite{IEEEhowto:13} inspected secure communication transmission in IRS assisted NOMA system for a practical eavesdropper scenario having imperfect channel state information. They proposed a joint IRS phase shift and transmit beamforming scheme to ensure secure transmission via IRS. Moreover, they used the alternating optimization algorithm which is a mathematical method to find out stationary point solutions. Further, in \cite{IEEEhowto:14} IRS assisted downlink NOMA system was considered with the aim of enhancing the performance of the rate and maximizing the signal to interference plus noise ratio (SINR). The authors achieved this by jointly optimizing the phase shift at the IRS and the transmit beamforming at the BS. They used efficient mathematical method based on semi-definite relaxation and block coordinated decent techniques. mmWave technology was included in the IRS NOMA scenario in \cite{IEEEhowto:15}, where the authors inspected the downlink scenario of IRS mmWave NOMA, and the formulated problem included joint optimization of the passive and active beamforming as well as power allocation with the aim of enhancing the performance of the system. To solve the problem, the authors used a mathematical method based on an iterative algorithm utilizing successive convex approximation and alternating optimization.

Nonetheless, the above studies \cite{IEEEhowto:11} - \cite{IEEEhowto:15} assumed that the channels between the IRS and users are known.  Such assumption contradicts the practical case, where IRSs are passive elements incapable of estimating channels.  In the case where part of the channels are unknown, the use of machine learning techniques can add value to the problem.



\subsection{Contributions}

In our research work, we address the aforementioned gap in surveyed literature by leveraging reinforcement learning (RL); in particular, deep reinforcement learning (DRL), to optimize the sum rate of a NOMA downlink system utilizing IRS under the assumption of unknown channel states information between the IRS and users. In particular, we exploit Deep Deterministic Policy Gradient (DDPG) due to its suitability for our scenario because the problem is non-convex, the objective function is non-convex and the unit modulus constraints are fundamentally non-convex. Therefore, it is an NP-hard problem \cite{IEEEhowto:8}, \cite{IEEEhowto:16}. The major challenge presents in the constant modulus constraint of the phase shift since the IRS can reflect the signal without amplifying it. Hence, it is not easy to obtain an optimal solution in closed form. The  use  of  DDPG  is very efficient   when   coping   with  intractable problems, it will  remove the  need  for  gathering a large dataset for training. DDPG method is anticipated to provide a solid and robust performance.

Our contributions in this work can be summarized as follows:

\begin{itemize}

\item First, we formulate the IRS NOMA downlink phase shift optimization problem with the objective of maximizing the sum-rate for NOMA users taking into consideration that the instantaneous channel states information between the IRS and users are unknown.

\item Second, we incorporate imperfect interference cancellation in practical NOMA within our system model formulation.

\item Third, DDPG based solution is proposed for predicting the best phase shift matrix in which the IRS learns the best way for reflecting the incident signals by modifying the phase.    

\item Fourth, numerical results reveal the effectiveness of the DDPG algorithm, since the sum rate value for the DDPG based IRS assisted NOMA system outperforms OMA based systems with minimum training overhead.

\end{itemize}

\subsection{Paper organization}
The rest of the paper is structured as follows. Section II outlines the system and channel model. Section III explains the DDPG-based phase control for IRS. Section IV presents the numerical results, and Section V concludes the paper.
 
\section{System Model}

\begin{figure}[t]
    \centering
    \includegraphics[width=0.7\linewidth]{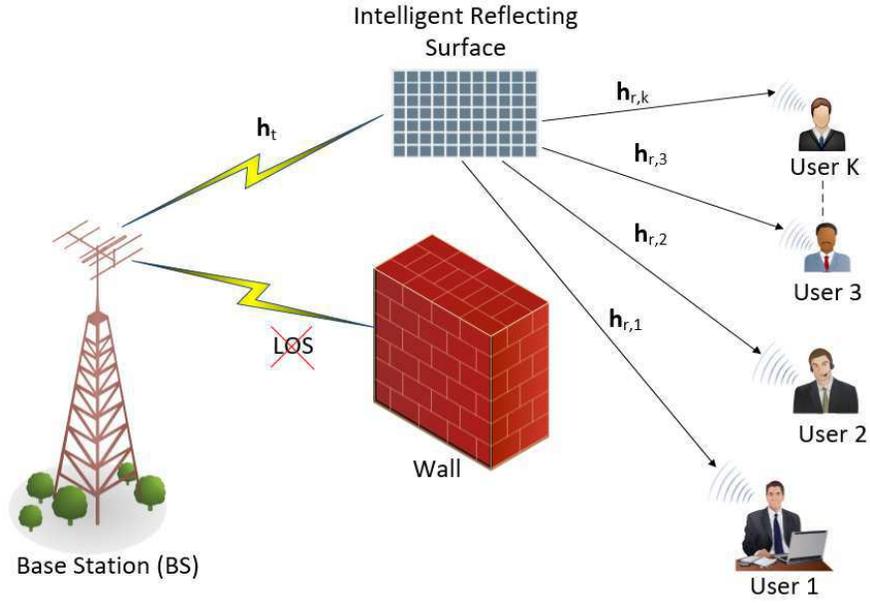}
    \caption{IRS Assisted Downlink NOMA}
     \label{fig:IRS Assisted Downlink NOMA}
\end{figure}

We consider the downlink of an IRS assisted NOMA system with $K$ users as shown in Fig.~\ref{fig:IRS Assisted Downlink NOMA}. Without loss of generality, the users are ordered according to their distance from the IRS such that user $1$ is the farthest user from the IRS and user $k$ is the nearest user to the IRS. Consequently, the users can be considered as ordered based on the expected value of their channel gains assuming  $|\mathbf{h}_{r,1}|^2 < |\mathbf{h}_{r,2}|^2<\ldots<|\mathbf{h}_{r,k}|^2$. All users as well as the BS are assumed to have a single antenna each. It is assumed that there is no direct line of sight (LOS) link between the users and the BS. Thus, the communication between the BS and users is performed through the IRS which is deployed with $M = M_x M_y$ reflecting elements, where $M_x$ and $M_y$ represent the number of passive elements in the IRS in every row and column, respectively.

The transmitted signal at the BS is 
\begin{align} x = \sum_{k=1}^{K} \sqrt{P_k} s_k, \end{align} where $s_k$ represents the signal for user $k$ with unit power (i.e., $\mathbb{E}[|s_k|^2] = 1$, $k \in \{1,...,K\}$, where $\mathbb{E}[.]$ denotes the expectation). The power allocated to each user is $P_k = \beta_k P$, where $\pmb{\beta} = [\beta_1, \beta_2, ..., \beta_k] $ is the vector of coefficients of the users' power allocated such that $\beta_1 + \beta_2 + ... + \beta_k = 1$ and $P$ is the BS total transmit power. The power allocated for users is fixed and it satisfies the relationship: $ P_1 > P_2 > ... > P_K $. When allocating power to users, we will take into consideration the following conditions $ P_1 > P_2 + P_3 + ... + P_K $, $ P_2 > P_3 + P_4 + ... + P_K $, and so on.

The received signal for user k can be written as:

\begin{align}
y_k & = \mathbf{h}_{r,k}^H \mathbf{\Phi} \mathbf{h}_{t} x + n,\\ 
\label{eq: Received Signal at user K}
y_k & = \mathbf{h}_{r,k}^H \mathbf{\Phi} \mathbf{h}_{t} \sum_{k=1}^{K} \sqrt{P_k} s_k + n,\nonumber
\end{align}

where $ \mathbf{h_t} \in \mathbb{C}^{M \times 1} $ is the channel between the BS and the IRS, and $  \mathbf{h}_{r,k} \in \mathbb{C}^{M \times 1} $ is the channel between the IRS and users. Both channels follow the Rician fading model:

\begin{align}
\mathbf{h_t} = \sqrt{\frac{K_1}{K_1+1}} \mathbf{\Bar ht} + \sqrt{\frac{1}{K_1+1}} \mathbf{\tilde ht},
\end{align}

\begin{align}
\mathbf{h_{r,k}} = \sqrt{\frac{K_2}{K_2+1}} \mathbf{\Bar h_{r,k}} + \sqrt{\frac{1}{K_2+1}} \mathbf{\tilde h_{r,k}},
\end{align}

where $K_1$ is the rician factor of $ \mathbf{h_t}$, $ \mathbf{\Bar ht} \in {C}^{M \times 1} $ and $ \mathbf{\tilde ht} \in {C}^{M \times 1} $ are the LoS component and non-LoS (NLoS) component, respectively. Similarly, $K_2$ is the rician factor of $ \mathbf{h_{r,k}}$, $ \mathbf{\Bar h_{r,k}} \in {C}^{M \times 1} $ and $ \mathbf{\tilde h_{r,k}} \in {C}^{M \times 1} $ are the LoS component and non-LoS (NLoS) component, respectively. The phase shift reflection matrix is represented by $ \large  \mathbf{\Phi} = \text{diag}(e^{j\theta_1},e^{j\theta_2},...,e^{j\theta_M}) $, and satisfies the constant modulus constraint $ |\phi_i|^2 = |e^{j\theta_i}|^2 = 1 $,  $ \forall i \in  \{1,2,...,M\}  $, because the IRS reflects the signal without amplifying it, where $\text{diag}(.)$ denotes a diagonal matrix. Further, the phase shift of the $ i^{th} $ passive reflecting element is denoted by $ \theta_i $, where the value of $ \theta_i $ is between $0$ and 
$ 2\pi $, and $ n  \sim \mathcal{C} \mathcal{N} (0,\sigma^2) $ represents the additive white Gaussian noise (AWGN).  

Hence, the received SINR at user $k$ can be represented by the following equation:

\begin{align}  \large \gamma_k = \bigg(\frac{|\mathbf{h}_{r,k}^H \mathbf{\Phi} \mathbf{h}_{t}|^2 P_k}{\sum_{i=k+1}^{K} |\mathbf{h}_{r,k}^H \mathbf{\Phi} \mathbf{h}_{t}|^2 P_i + \sigma ^2}\Bigg), \label{eq:Signal to Noise Ratio}
\end{align}  

when $k = K$, the term $  \large \sum_{i=k+1}^{K} |\mathbf{h}_{r,k}^H \mathbf{\Phi} \mathbf{h}_{t}|^2 P_i = 0 $.

Furthermore, the data rate of user $k$ is represented by: 
\begin{align} \large R_k = \log_2(1+\gamma_k). \end{align}  

Our objective in this study is to maximize the sum-rate of all users where:
\begin{align}   
\large
R_{sum} =\sum_{k=1}^{K} \log_{2} \left (1 + \large \gamma_k \right).\label{eq:SumRate}
\end{align}

Therefore, the formulated problem at the IRS is to obtain the phase shift reflection matrix $ \mathbf{\Phi} $ that maximizes $ R_{sum} $ for users.

\begin{align} 
\large
\max_{\Phi}\sum_{k=1}^{K} & \log_{2} \left (1 + \large \gamma_k \right), \label{eq:max_SumRate_without_SIC} \\
s.t.~&|\mathbf{\phi}_i|^2 = 1 ,   \forall i \in  \{1,2,...,M\} ,\nonumber\\
&\large {\beta_1 + \beta_2 + ... + \beta_k = 1}, \nonumber
\end{align}  

The above equation \eqref{eq:max_SumRate_without_SIC} is valid for perfect successive interference cancellation (SIC) which is the ideal case, where the interference from the far users is assumed to be perfectly eliminated at the near users receiver. In this case near users have perfect knowledge of the far users' data signal. Nonetheless, in the case of imperfect SIC, the interference of the far users is not perfectly removed at the near users receiver. In this case, the data signal of the far user is not perfectly known at the near user due to distortion caused by fading and AWGN which is more realistic practically. Thus, the received SINR in \eqref{eq:Signal to Noise Ratio} at user $k$ can be rewritten as:

\begin{align} 
\small \Tilde{\gamma}_k = \frac{|\mathbf{h}_{r,k}^H \mathbf{\Phi} \mathbf{h}_{t}|^2 P_k}{\large \epsilon \sum_{j=1}^{k-1} |\mathbf{h}_{r,k}^H \mathbf{\Phi} \mathbf{h}_{t}|^2 P_j+ \sum_{i=k+1}^{K} |\mathbf{h}_{r,k}^H \mathbf{\Phi} \mathbf{h}_{t}|^2 P_i + \sigma ^2}, 
\label{eq:SumRate_with_SIC}
\end{align}
when $k$ = 1, the term $  \large \sum_{i=1}^{k-1} |\mathbf{h}_{r,k}^H \mathbf{\Phi} \mathbf{h}_{t}|^2 P_j = 0 $. The term epsilon ($  \large \epsilon $) represents the fraction of the residual interference leftover due to imperfect SIC.

Accordingly, the optimization problem in \eqref{eq:max_SumRate_without_SIC} becomes 

\begin{align} 
\large
\max_{\Phi}\sum_{k=1}^{K} & \log_{2} \left (1 + \large \Tilde{\gamma}_k \right), \label{eq:max_SumRate_with_SIC} \\
s.t.~&|\mathbf{\phi}_i|^2 = 1 ,   \forall i \in  \{1,2,...,M\} ,\nonumber\\
&\large {\beta_1 + \beta_2 + ... + \beta_k = 1}, \nonumber
\end{align}  

\subsection{Upperbound on Performance}

To measure the performance of the DDPG algorithm and to verify that our sum-rate values approach the upperbound, an exhaustive search method is used to search for the optimum phase shift matrix that results in the maximum sum-rate as shown in Algorithm \ref{Alg:Exhaustive Search}. Further, to avoid the huge complexity of the exhaustive search scheme, we will consider a limited number of IRS reflecting elements as a case proof that our DDPG algorithm can track the upperbound. For every IRS element we will consider the phases between 0 and $2\pi$ with a step size of 16, this will give us $16^M$ combinations of phase shift matrices. Then we will calculate the sum-rates accordingly for $K$ users.

\begin{algorithm}[t]
\caption{Exhaustive Search for the Phase Shift Matrix}
    \label{Alg:Exhaustive Search}
    \small{
\begin{algorithmic}[1]
\State  Initialize M = 4, $\Delta\Phi  = \frac{2\pi}{16}$,
        \For {$\phi_1 = 0:\frac{2\pi}{16}:2\pi;$}
        \For {$\phi_2 = 0:\frac{2\pi}{16}:2\pi;$}
        \For {$\phi_2 = 0:\frac{2\pi}{16}:2\pi;$}
        \For {$\phi_3 = 0:\frac{2\pi}{16}:2\pi;$}
        
\State Calculate and store $R_{sum} (\phi_1, \phi_2, \phi_3, \phi_4)$ 
        
    \EndFor
    \State \textbf{end for}
    \EndFor
    \State \textbf{end for}
    \EndFor
    \State \textbf{end for}
     \EndFor
    \State \textbf{end for}
\State Find $\Phi^* =  \argmax_{\phi_1, \phi_2, \phi_3, \phi_4} R_{sum} $

\end{algorithmic}}
\end{algorithm}


\subsection {OMA Baseline Scheme}

The signal model for OMA is assumed such that the resources (frequency / time) are divided equally between the $K$ users. This enables OMA users to receive the signal with free interference, whereas the merit of NOMA is the simultaneous transmission and the interference can be controlled. However, to serve K OMA users, FDMA / TDMA requires $K$ time slots. The first user will use the first frequency/time slot, the second user will use the second frequency/time slot, and user $K$  will use the $K^{th}$ frequency/time slot accordingly.

The transmitted signal by the BS is given by:
\begin{align} x^{OMA}_k = \sqrt{P}  s_k, \end{align}
The received signal at the user side can be expressed as:
\begin{align} y^{OMA}_k = \sqrt{P} s_k \mathbf{h}_{r,k}^H \mathbf{\Phi} \mathbf{h}_{t} + n, \end{align}
Hence, the received SNR at user k can be represented as:
\begin{align} \gamma_k = \bigg(\frac{|\mathbf{h}_{r,k}^H \mathbf{\Phi} \mathbf{h}_{t}|^2 P}{\sigma ^2}\Bigg),  \end{align}  
Further, data rate of user k is represented by: 
\begin{align}  R^{OMA}_k = \frac{1}{K} log_2(1+\gamma^{OMA}_k), \end{align}  
Therefore, the sum-rate of OMA can be expressed as:
\begin{align} R^{OMA}_{sum} = \sum_{k=1}^{K}  R^{OMA}_k, \end{align} 
\begin{align}
R^{OMA}_{sum} = \frac{1}{K} \sum_{k=1}^{K} \log_{2} \Bigg(1+ \frac{|\mathbf{h}_{r,k}^H \mathbf{\Phi} \mathbf{h}_{t}|^2 P}{\sigma ^2}\Bigg).
\end{align}
\vspace {0.25 cm}

\section {Proposed DRL-Based Phase Control for IRS}
\subsection {Overview on DRL method:}
 
\begin{figure}[H]
    \centering
    \includegraphics[width=0.6\linewidth]{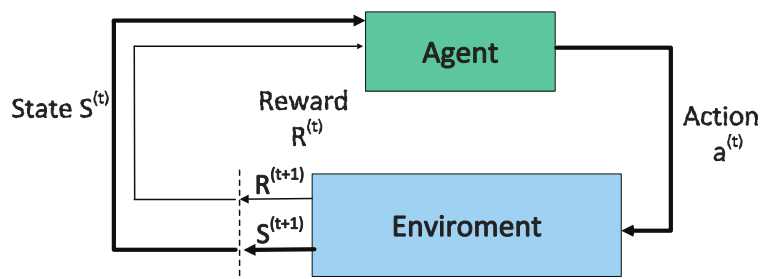}
    \caption{DRL Model}
    \label{fig:DRL Model}
\end{figure}

Reinforcement Learning (RL) method is a field of machine learning that allows the agent in an interactive environment to learn by trial and error relying on a feedback from its own experiences. The model in Fig.~\ref{fig:DRL Model} demonstrates the basic concept of the RL. Further, the fundamental factors that characterize the RL problem are the agent, environment, state, reward, policy, and value. The environment is the physical surrounding where the agent operates, the state is the agent's current status, the reward is the feedback the agent experience from the environment, the policy is the process of mapping the state of the agent to actions, and the value is the agent's future reward that it receives when taking action in a specific state. The agent at time $t$ gets state $s^{(t)}$ from the environment and selects action $ a^{(t)} $ based on policy $ \pi $. After selecting the action the state changes from $s^{(t)}$ to $ s^{(t+1)}$ and generate reward $ r^{(t)} $. The objective of RL is to maximize the total reward, the reward is the performance measure of a specific action during the current state. RL models the interaction between the agent and the environment as Markov's decision process (MDP). Many RL algorithms use dynamic programming methods, but the prime difference between RL algorithms and dynamic programming is that RL algorithms don't assume the awareness of a specific MDP's mathematical model. Further, RL algorithms are aimed for large MDPs in which specific procedures become impracticable \cite{IEEEhowto:17}. 

Q-learning (QL) is a model-free RL method since it does not require a model, it obtains the best action under the current state, where "Q" denotes the quality. Further, it is an off-policy algorithm since QL learns actions that are outside the current policy so it may take actions that are random and therefore the policy is not needed here. When the agent interacts with the environment and needs to update the action-state pairs in the Q-table, it has two choices, either to explore or exploit. Exploration is to explore the environment in order to obtain information and knowledge about it, in this case the agent acts randomly. Whereas exploitation is to exploit the available "already known" information about the environment so as to maximize the reward. Precisely, QL selects the optimal policy in order to maximize the total reward. 

\vspace {0.25 cm}

The trade-off between exploration and exploitation is balanced by using a specific parameter  $\zeta$ which can be set based on how often we want to exploit or explore. Moreover, QL is based on the concept of a Q-function, which is a function of states and actions. The Q-function $ Q^{\pi}(s,a) $ measures the expected  discounted sum of rewards or the return achieved from selecting action $ a $ in a state $ s $ based on policy $ \pi $. It evaluates how good a particular action is, for a given state. The optimal Q-value $ Q^*(s,a) $  is defined as the maximum return achieved from a given state $ s $ and action $ a $ while obeying the optimal policy. In order to update the Q-value for any action we use the Bellman equation. It provides us the best reward and the optimal policy to achieve this reward. It means that the maximum value of the return from the action and state is equal to the expectation of the current maximum possible reward plus the maximum possible long-term reward achieved from the next state s' discounted by $ \gamma \in [0,1] $ which is the discount factor, and the equation is expressed as: 


\vspace {0.25 cm}
\begin{align} Q^*(s,a) = \mathbb{E} [r (s,a) +\gamma \max_{a'} Q^*(s',a')], \end{align}

\vspace {0.25 cm}

\noindent where $ \mathbb{E} [.] $ denotes the expectation. The updates will occur after each action and it will end when the episode finishes. In order to converge and learn optimal values, the agent needs to learn and explore many episodes. Furthermore, there are three important steps in the update, the first step is that the agent will take action for each state and receive a reward. The second step is that the agent will either select the action by checking the highest value in the Q-table or by random $ \zeta $. The third step is to update the Q-values employing the following formula:

\vspace {0.25 cm}
\begin{align} Q^{new}(s,a) = (1-\alpha) Q(s,a) + \alpha [r (s,a) +\gamma \max_{a'} Q(s',a')], \end{align}
\vspace {0.25 cm}

\noindent The step size is adjusted using the learning rate $ \alpha $  which measures the acceptance of the new value compared to the old one \cite{IEEEhowto:17}.

 \begin{figure} [H]
    \centering
    \includegraphics[width=0.5\linewidth]{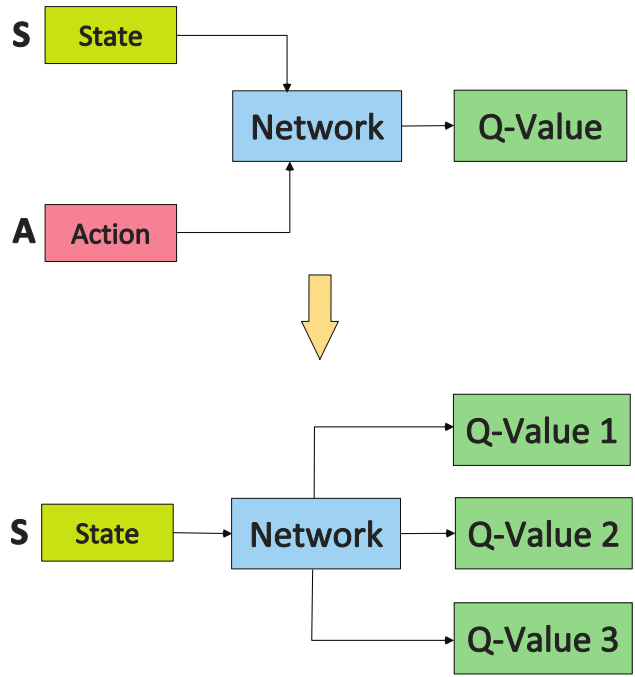}
    \caption{Q-Learning vs Deep Q-Learning}
    \label{fig:Q-Learning vs Deep Q-Learning}
\end{figure}

\vspace {0.25 cm} 

Deep Q-Learning (DQL) uses  neural network that approximates the Q-function. Fig.~\ref{fig:Q-Learning vs Deep Q-Learning} shows a comparison between QL and DQL, instead of calculating Q-values immediately for each action and state via iterations, we will utilize a function approximator for estimating Q-values for each possible action. To perform this we need to employ neural networks (NN). The NN accepts states as input from the environment and generates the estimated Q-values for every action the agent can select. Furthermore, in many problems in DQL it is not practical to represent the Q-function based on $ s $ and $ a $ only, the reason is that we will train the neural network (NN) with parameters $ \theta $ to evaluate the Q-values such that $ Q(s,a,\theta) \approx Q^*(s,a) $.

\vspace {0.25 cm}
\begin{align} Q_\pi(s,a;\theta) = \mathbb{E_\pi} [R^{(t)}_{s^{(t)}=s,a^{(t)}=a}], \label{eq:Maximize_Q-Value} \end{align}

\vspace {0.25 cm}
\noindent where $ R^{(t)}= \sum_{t=0}^{+\infty} \gamma^t r^{(t)} $ is the cumulative expected reward. The aim of the DQN is to maximize the value of the Q-function in \eqref{eq:Maximize_Q-Value} by DNN training. The optimal Q-value is obtained from the Bellman equation. At this stage, we can calculate the loss by comparing the Q-value for a given state-action pair to the target value which is the right hand side of the Bellman equation expression which gives us the below expressions:

\begin{align} \text{Loss = } Q^*(s,a;\theta) - Q(s,a;\theta), \end{align}

\begin{align}  \mathbb{E} [r (s,a;\theta) +\gamma \max_{a'} Q^*(s',a';\theta)] - \mathbb{E} [\sum_{t=0}^{+\infty} \gamma^t r_t], \end{align}

\noindent Next the NN will update its weight values in the policy network based on the gradient descent and back propagation algorithms. We will keep repeating this process time after time for many episodes until we sufficiently minimize the loss \cite{IEEEhowto:17}. 


\vspace {0.25 cm}
Policy gradient (PG) is  on-policy, model-free RL method, used for continuous action space. The PG agent is a RL agent which is policy-based, it aims directly to maximize the expected reward by obtaining a parametrized policy that generates a trajectory $ \tau $, where the trajectory represents the states, actions and rewards i.e. $ s_0,a_0,r_1,s_1,a_1,r_2,... $. In other words, we need to find out the parameters $\theta $ that maximizes $J$, where $ \theta$ represents the weights of the neural networks.

\begin{align} J(\theta) = \mathbb{E_\pi}[r(\tau)], \end{align}

\noindent where $ r(\tau) $ denotes the total reward for a specified trajectory $ \tau $. A well known approach in machine learning to solve the maximization problem is the gradient descent or ascent to step through the parameters. By using the gradient ascent we get the following equation:

\begin{align} \theta_{t+1} = \theta_t + \alpha \nabla J(\theta_t), \end{align}  

\noindent where $\alpha $ represents the learning rate.

\begin{align} \nabla J(\theta) = \nabla \mathbb{E_\pi}[r(\tau)] = \mathbb{E_\pi}[r(\tau) \nabla \log (\pi(\tau))], \end{align}

By reformulating the gradient we get:\\

\noindent $\nabla \mathbb{E}_{\pi_{\theta_t}}[r(\tau)] =  \nabla \int \pi (\tau) r(\tau) d\tau , $\\
\vspace {0.25 cm}
$ \nabla \mathbb{E}_{\pi_{\theta_t}}[r(\tau)] =  \int \nabla \pi (\tau) r(\tau) d\tau , $\\
\vspace {0.25 cm}
$ \pi (\tau) \nabla log \pi(\tau) =  \frac{\pi (\tau) \nabla \pi(\tau)}{\pi (\tau)} = \nabla \pi (\tau) $, \\
\vspace {0.25 cm}
$\nabla \mathbb{E}_{\pi_{\theta_t}}[r(\tau)] =  \int \pi (\tau) \nabla log \pi(\tau) r(\tau) d\tau , $ \\

The theory of the policy gradient states that the derivative of the expectation of the reward is equivalent to the expectation of the reward times the gradient of the log policy $ \pi_\theta $ :

\begin{align} \nabla \mathbb{E}_{\pi_{\theta_t}}[r(\tau)] = \mathbb{E}_{\pi_{\theta_t}} [r(\tau) \nabla_{\theta_t} \log (\pi_{\theta}(\tau))], \end{align}

Therefore, the policy training is demonstrated as a gradient ascent process:

\begin{align} \theta_{t+1} = \theta_t + \alpha \mathbb{E}_{\pi_{\theta_t}} [r(\tau) \nabla_{\theta_t} \log (\pi_{\theta}(\tau))], \end{align}

\noindent where \begin{math} Q_{\pi_{\theta_t}}(\tau) \end{math} is the Q-value for the trajectory \begin{math} \tau \end{math}, and policy \begin{math} \pi_{\theta_t} \end{math}. Throughout the training, the PG agent estimates the probability of selecting each action and chooses actions by random depending on the probability distribution. Before it learns from experience and updates the policy parameters, the PG agent performs a full training episode utilizing the existing policy. The disadvantage of PG method is that the network policy is updated after the completion of the episode only. This slackens the convergence rate \cite{IEEEhowto:17}.  

\subsection {DDPG-based IRS phase control method}

\begin{figure}[t]
    \centering
    \includegraphics[width=0.5\linewidth]{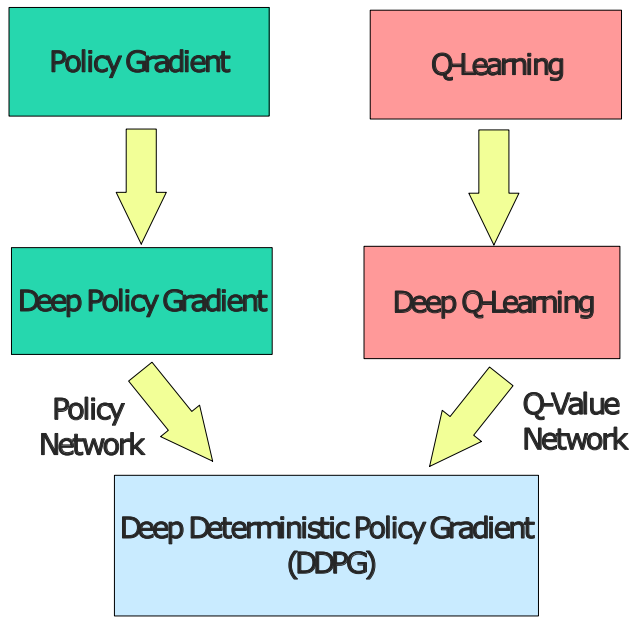}
    \caption{DDPG merges both DPG and DQN}
    \label{fig:DDPG merges both DPG and DQN}
\end{figure}

We propose a DDPG-based IRS phase control method considering the optimization problem in \eqref{eq:max_SumRate_without_SIC}. Deep Q-Networks are not suitable because they deal with discrete time spaces only. Moreover, the convergence of the policy gradient (PG) algorithm is not sufficient in the context of wireless communication. DDPG merges the Q-networks and the PG scheme as shown in Fig.~\ref{fig:DDPG merges both DPG and DQN}, and overcomes the disadvantages of both algorithms\cite{IEEEhowto:17}.  

\begin{figure} [t]
   \centering
    \includegraphics[width=1\linewidth]{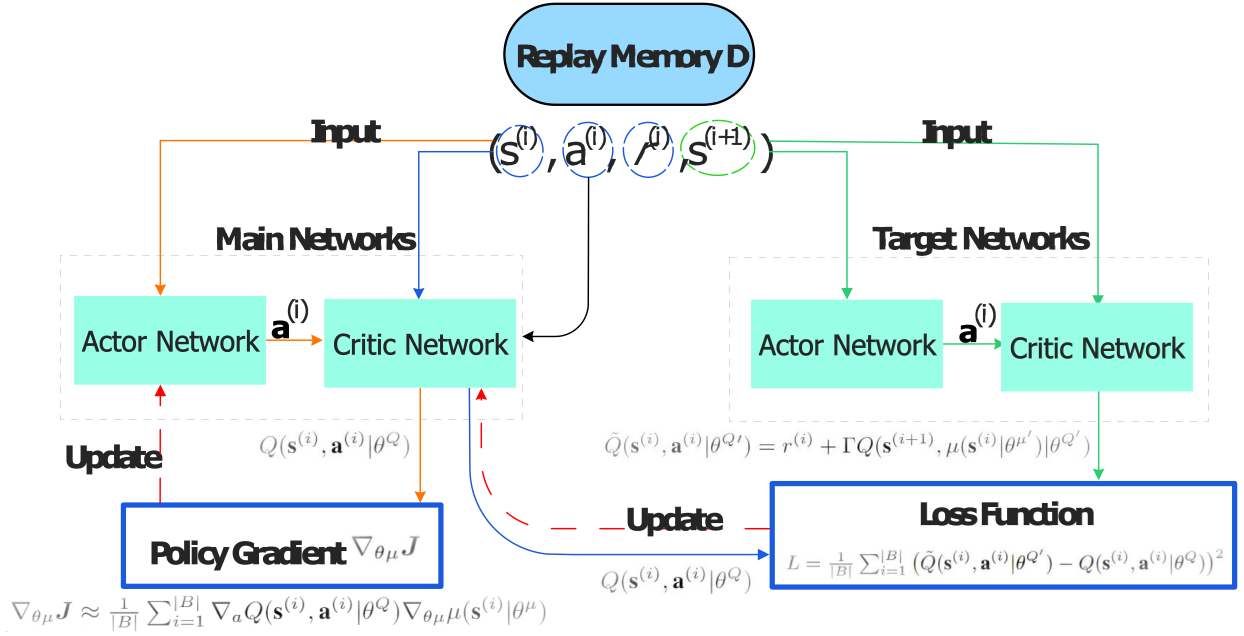} 
     \caption {DDPG Model}  
     \label{fig:DDPG Model}
\end{figure}

DDPG is a model-free reinforcement learning technique which combines the advantages of policy gradients and Q-learning. DDPG uses the Bellman equation and the off-policy data to learn the Q-function, and then uses the Q-function to learn the policy. DDPG consists of four neural networks (see Fig.~\ref{fig:DDPG Model}); one for actor network, one for critic network, one for target actor network, and one for target critic network, which ensures the stability. The optimization problem in \eqref{eq:max_SumRate_without_SIC} can be solved using DDPG by learning the policy.

\vspace {0.25 cm}
The actor network is a policy network that accepts state as input and generates the precise action continuously. In deep Q-networks the optimal action is obtained by calculating the argmax over all the Q-values for a finite number of discrete actions. In DDPG the actor network performs the same but for continuous action spaces, it generates the actions directly by taking the argmax, and selects actions $ a = \mu (s|\theta^\mu) $ from a continuous action settings $\mathcal{A} $, where  $ \mu $ is the policy, $s$  denotes the states, and $ \theta^\mu $ denotes the parameters of the deterministic policy network (DPN). Thus, the actor is a DPN that calculates the action directly, rather than generating the probability distributions over all actions. The critic network accepts states and actions as input and generates the Q-value, so it is considered as the Q-value network $ Q(s, a | \theta^{Q}) $, where $ \theta^Q $ represents the parameters of the Q-network. The critic network evaluates the performance of the selected action. Therefore, DDPG is an enhancement for the actor-critic vanilla network, its aim is to maximize the Q-value which is the output, and it can only be utilized for environments having continuous action settings. The optimal action is expressed as:

\begin{align} \mu^* (s|\theta^{\mu}) = \argmax_a Q^*(s,a|\theta^{Q}) \end{align}

\noindent where $ Q^*(s,a|\theta^{Q}) $ is the optimal Q-value function. Further, to maximize the Q-value a replay memory $\mathcal{D}$ is utilized to minimize the correlation of various training samples. This is significant for the algorithm behavior in order to be stable. The replay memory needs to be adequately large to include a broad range of previous experiences. Moreover, DDPG makes use of target networks to increase the stability during the training. 

A copy from the actor and critic network are formed  to find out the Q-value for the next state i.e. $ a = \mu' (s| \theta^{\mu'}) $ and $ Q'(s, a| \theta^{Q'}) $. Based on the main networks, the weights of these target networks are updated periodically. In deep Q-networks, the weight of the main network is copied periodically to the target network and this called "hard update", whereas in DDPG "soft update" is performed where a only fraction of the weights of the main network are transferred to the target network as expressed below:

\begin{align} 
\theta^{Q'}\leftarrow\tau\theta^Q+(1-\tau)\theta^{Q'}, \label{eq:Updating Target Wieghts} \\
\theta^{\mu'}\leftarrow\tau\theta^\mu+(1-\tau)\theta^{\mu'},
\end{align} 

\vspace {0.25 cm}

\vspace {0.25 cm}

\newpage

\begin{algorithm}[t]
\caption{DDPG-based IRS Phase Control Training}
    \label{Alg:DDPG}
    \small{
\begin{algorithmic}[1]
\State \textbf{Initialization:} Set $t=0$ and initialize reply buffer of DDPG agent $\mathcal{D}$ with capacity M.
\State Randomly initializes the weights of actor networks $\theta^{\mu}$ and critic networks $\theta^Q$.
\State Initialize target networks: $\theta^{\mu'}\leftarrow\theta^{\mu}$ and $\theta^{Q'}\leftarrow\theta^Q$. 
\For{$t=1$ to $\infty$}
\State Observe state $\textbf{s}^{(t)}$ and and select an action with exploration OU noise $\textbf{a}^{(t)}=\mu(\textbf{s}^{(t)}|\theta^\mu)+\textbf{n}_t$
\State Execute action $\textbf{a}^{(t)}$ at IRS.  
\State Receive the immediate reward $r^{(t)}$, and observe next state $s^{(t+1)}$, store transition $(\textbf{s}^{(t)},\textbf{a}^{(t)},r^{(t)},\textbf{s}^{(t+1)})$ in $D$.
\State Randomly sample mini-batch transitions from $\mathcal{D}$:

$B\leftarrow\{(\textbf{s}^{(i)},\textbf{a}^{(i)},r^{(i)},\textbf{s}^{(i+1)})\} \in \mathcal{D}$ .
\State Compute the targets:

$\Tilde{Q}(\textbf{s}^{(i)},\textbf{a}^{(i)}|\theta^{Q'})=r^{(i)}+\Gamma Q(\textbf{s}^{(i+1)},\mu(\textbf{s}^{(i)}|\theta^{\mu'})|\theta^{Q'})$
\State Update the $\theta^Q$ in critic network by minimizing the loss:

$L = \frac{1}{|B|}\sum_{i=1}^{|B|}\big(\Tilde{Q}(\textbf{s}^{(i)},\textbf{a}^{(i)}|\theta^{Q'})-Q(\textbf{s}^{(i)},\textbf{a}^{(i)}|\theta^Q)\big)^2$
\State Update the $\theta^\mu$ in actor network according to the sampled policy gradient:

$\nabla_{\theta\mu}\boldsymbol{J}\approx \frac{1}{|B|}\sum_{i=1}^{|B|}\nabla_a{Q}(\textbf{s}^{(i)},\textbf{a}^{(i)}|\theta^{Q})\nabla_{\theta\mu}\mu(\textbf{s}^{(i)}|\theta^\mu)$

\State Update the target networks:

$\theta^{Q'}\leftarrow\tau\theta^Q+(1-\tau)\theta^{Q'}$

$\theta^{\mu'}\leftarrow\tau\theta^\mu+(1-\tau)\theta^{\mu'}$
    \EndFor
    \State \textbf{end for}
\end{algorithmic}}
\end{algorithm}

\noindent where $\tau \ll 1 $. Soft updates are significant in order to accelerate the convergence of the Actor-Critic process since it stabilizes learning. The main networks which are copied are called evaluation networks. The target networks and the evaluation networks have the same structure but the difference is in parameters. The target networks are delayed networks compared to the main networks \cite{IEEEhowto:17}. 

\subsection{RL System Mapping}
The first step in solving a problem using RL is to map the problem into the key components of an RL system; namely, state-space, action space, and reward function.  In the following, we discuss this mapping as well as the general behaviour of the RL method using DDPG. 

\subsubsection{State-space}
The state space of the DDPG agent at timestep ${(t)}$ can be defined as follows
\begin{align}
    \textbf{s}^{(t)}=[~\textbf{h}_t^{(t)},~\boldsymbol{\Phi}^{(t-1)},~\boldsymbol{\hat{\gamma}}^{(t-1)}~],
\end{align}
where $\textbf{h}_t$ represents the channel gain between the source and IRS, $\boldsymbol{\Phi}$ is the last phase action taken by IRS, and $\boldsymbol{\hat{\gamma}}$ is the estimated SINR values of the users based on their data rates for that action (i.e., $\boldsymbol{\gamma}=[\gamma_1,\gamma_2,\dots,\gamma_k,\dots,\gamma_K]$ ).
\subsubsection{Action-space}
The action space definition is defined by the policy function as follows
\begin{align}
    \textbf{a}^{(t)}=\mu(\textbf{s}^{(t)}|\theta^\mu)+\textbf{n}{(t)}
\end{align}
where $\mu$ is the policy function and $\theta^\mu$ is parameters (i.e., weights of neural network), and $\textbf{n}(t)$ is the Ornstein-Uhlenbeck (OU) process-based action noise~\cite{IEEEhowto:18}.
Final output is an array that defines the phase of each element in the IRS.
\subsubsection{Reward function}
The reward function is defined based on the current channel capacity and the maximum capacity ever reached as follows:
\begin{align}
    r^{(t)} = R_{sum}^{(t)} - R_{sum,max},
\end{align}
where $R_{sum}^{(t)}$ is the actual sum-rate of the users, while $R_{sum,max}$ is the maximum sum-rate achieved.
\subsubsection{Exploration vs. Exploitation}
Since the action space of the DDPG is continuous, the exploration of action space is handled with noise generated by the OU process.
OU process samples noise from a correlated normal distribution.
\subsubsection{DDPG Algorithm}
As shown in Algorithm~\ref{Alg:DDPG}, we begin initializing the replay buffer $D$ of the agent with transaction capacity $M$ in step~1.
In step~2, we initialize the weights of actor and critic networks for the agent.
The target networks are initialized by copying the same weights in step~3.
From step 4 to 13, represents each iteration (i.e., timestep $t$).
In each iteration (i.e., timestep $t$), we observe the state $\textbf{s}$ for the agent (IRS), determine an action (i.e., phase value) with exploration noise based on OU process in step~5.
After the agent determined and executed the action, a reward $r^{(t)}$ is received and new state $\textbf{s}^{(t+1)}$ is observed, and transactions are stored in respective replay memories in steps~6~and~7.
A random mini-batch of transitions are sampled in step~8.
Using the Bellman equation, the actors and critic networks' targets are computed in step~9.
The critic network weights are updated by minimizing the loss using computed targets in step~10.
The actor network weights are updated for the sampled policy gradient in step~11.
Finally, the agent's target networks are updated using the update rate ($\tau$) for stability in step 12.

\subsubsection{Neural Network Architecture}

DDPG agent's architecture consists of $4$ neural networks including actor and critic networks and target actor and critic for stability.
Both actor and critic networks consist of $2$ hidden layers, with $256$ hidden nodes in each layer.
The actor-network input has the size of $2M+K$, and output is $M$, thanks to the continuous definition of DDPG.
As can be seen from DDPG agents' structure, it allows scalability to a much larger extent with linearly increasing complexity.

\subsection{Discussion on Complexity} To reveal the value of using DDPG, we provide a quick quantitative analysis of the exhaustive search algorithm complexity, $\mathcal{N}_E$ versus the complexity, $\mathcal{N_D}$ of the proposed DDPG based algorithm.  The complexities can be easily deduced from the description of the algorithms given in Algorithm~\ref{Alg:Exhaustive Search} and Algorithm~\ref{Alg:DDPG}.  For exhaustive search asssuming $K$ users, $M$ IRS elements and $N = \frac{2\pi}{\Delta \Phi}$ phase change steps, we can write the complexity as 
\begin{align}
\mathcal{N}_E=O(K\times N^M).
\end{align} 

For the DDPG based system, the complexity for the trained network (steady state complexity) depends mainly on the forward network architecture (Actor Network).  Assume the number of states (size of actor network input) is $S$, number of hidden layers is $n$, number of neurons in each hidden layer is $U$, number of actions (i.e., phase of each IRS element) which is size of the output layer is $A$, and the DDPG algorithm will always provide the action of the highest reward for the $A$ distinct actions as output.
Therefore, the complexity of the DDPG can be written as
\begin{align}
    \mathcal{N}_D=O(S\times n \times U \times A).
\end{align} 

Thus, the complexity of DDPG is much lower than that of the exhaustive search as the number of users or the number of IRS elements increases.

\section{Numerical Results}

\begin{figure}[t]
    \centering
    \includegraphics[width=0.6\linewidth]{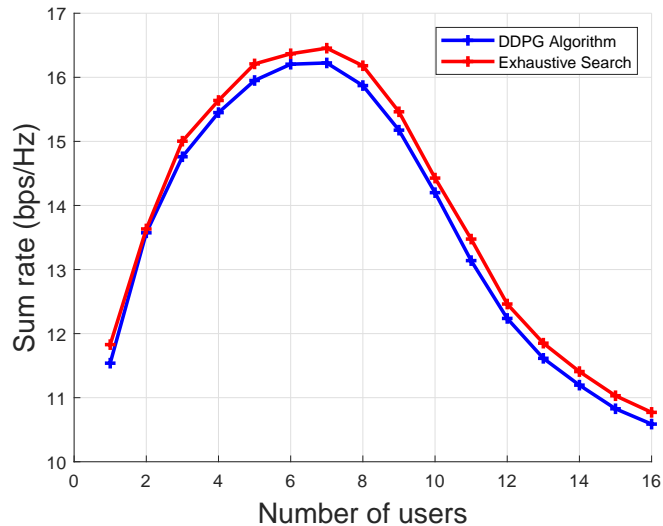}
    \caption{Upperbound on Performance vs Proposed DDPG Algorithm. M = 4, K = 16, and $\Delta\Phi  = \frac{2\pi}{16}$.} 
    \label{fig:Upperbound on Performance}
\end{figure}

In this section, we first measured the performance of the DDPG algorithm to make sure that the sum-rate values calculated are close to the upperbound. By using the exhaustive search algorithm we calculated the maximum sum-rate by obtaining the optimum phase shift matrix and assuming that the channel is known. The exhaustive search scheme is very complex, so the number of IRS reflecting element used is $M$ = 4 rather than $M$ = 16. This is to verify that our DDPG algorithm can approach the upperbound. For each element we considered the phases between 0 and $2\pi$ with a step size of 16. Thus, the total number of combinations of phase shift matrices is $16^4$.  The sum-rates are calculated for 16 users and for monte-carlo simulations equal to 1000. Fig.~\ref{fig:Upperbound on Performance} reveals that NOMA sum-rate generated by the DDPG algorithm approaches the upperbound and it is close to optimal. The complexity of the exhaustive search algorithm can be calculated as 1000 x $16^5$ which is equal to  $1.0486 \times 10^9$ iterations with elapsed time equal to 12.51 hours. 

 \begin{figure}[t]
    \centering
    \includegraphics[width=0.6\linewidth]{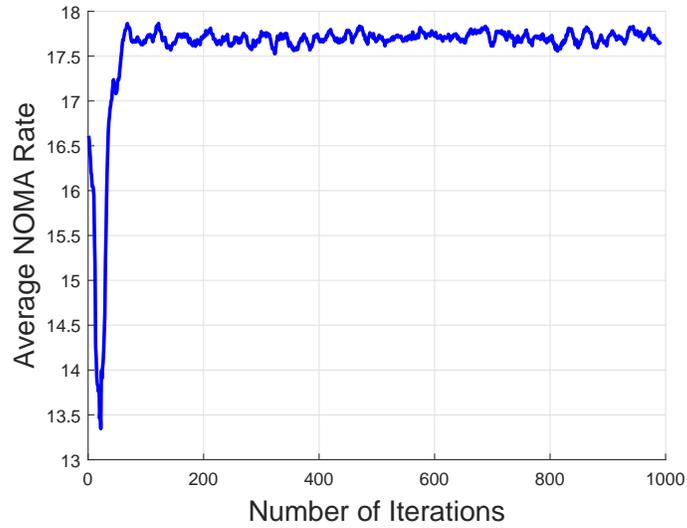}
    \caption{NOMA Sum Rate vs Iteration Plots}
    \label{fig:NOMA Sum Rate vs Iteration Plots}
\end{figure}


Moreover, the result in Fig.~\ref{fig:NOMA Sum Rate vs Iteration Plots} verifies the convergence of our DRL algorithm. It shows the average NOMA rate verses the iteration plots. The average rate is increasing with time. This means that the training process is conducted successfully. Further, the simulation results below reveal the performance of our DRL based IRS NOMA system with IRS reflecting elements $ M = 16 $. The default parameters used in the simulation are shown in Table \ref{table:2}. The number of of users is $ K = 32 $, the number of BS antennas is $ N_t = 1 $, the number of antennas per each user is $ N_r = 1 $, the distance between the BS and the IRS is 50 m and the distances between the IRS and the users are randomly generated between 200 and 1500 m. The channel between the BS and the IRS and the channel between the IRS and users follow the rician fading model with rician factor K1 = K2 = 10. However, the channel between the BS and the IRS is assumed to be perfectly estimated, whereas the channels between the IRS and users are assumed to be unknown. The bandwidth is 10 MHz, the BS transmit power Pt is 40 dBm, and the noise power spectral density equals -174 dBm/Hz. Simulation results are generated using $ 10^3 $ monte-carlo runs.

 \begin{figure}[t]
    \centering
    \includegraphics[width=0.6\linewidth]{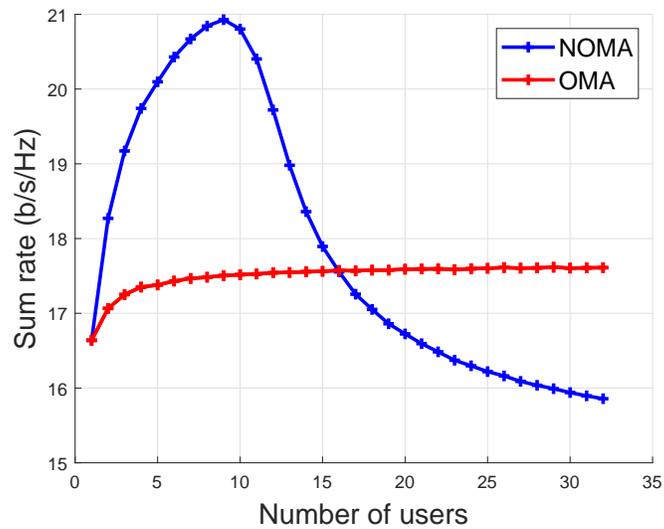}
    \caption{Comparison between NOMA and OMA sum rates}
    \label{fig:Comparison between NOMA and OMA sum rates}
\end{figure}

\begin{table} [t]
\caption{Parameters Used in Simulation} 
\centering 
\begin{tabular}{| c | c |} 
\hline 
\textbf{Simulation Parameters} & \textbf{Values}\\ 
\hline 
Number of Users (K) & 32 \\ 
\hline
Number of Reflecting Elements (M) & 16 \\
\hline
Number of BS antennas $ N_t $ & 1 \\
\hline
Number of antennas per user $ N_r $ & 1 \\
\hline
Distance between BS and IRS & 50 \\
\hline
Distance between the IRS and the users & 200 - 1500 \\
\hline
BS transmit power & 40 dBm \\  
\hline 
Bandwidth & 10 MHz \\
\hline
Noise power spectral density & -174 dBm/Hz\\  
\hline
BS to IRS Path loss exponent & 2\\ 
\hline
IRS to users Path loss exponent & 2.8\\ 
\hline
Rician Factor & 10\\  
\hline
Critic learning rate & 0.001 \\
\hline
Actor learning rate & 0.0005 \\
\hline
Discount factor $\Gamma$ & 0.05 \\
\hline
Coefficient of Soft Updates $\tau$ & 0.05 \\
\hline
Batch size & 64 \\  
\hline
Buffer Capacity $ \mathcal{C} $ \ & 10000\\ 
\hline
\end{tabular}
\label{table:2} 
\end{table} 

In the proposed DDPG algorithm, the actor and critic networks are both dense neural networks (DNN). The input of the actor network is the number of states that contains 128 neurons while the output is the number of actions which contains 16 neurons. The hidden layers in the actor network are two layers that contain 256 neurons each, followed by ReLU activation function. The output layer of the actor network uses the tanh(·) function in order to provide enough gradient. For the critic network, the input layer is the number of states and the number of actions. The state input is followed by two dense layers of 128, and 256 neurons respectively with ReLU activation functions, and the action input is followed by one dense layer of 128 neuron. Both outputs are passed via separate layer before concatenating to represent the input of the critic network. After that, two hidden layers are added each of 256 neurons with ReLU activation functions. This is pursued by the output layer of the critic network which contains 16 neurons. Both actor and critic main networks use Adam optimizer to update parameters. Moreover, we set the number of steps in each episode B = 1000, the actor learning rate = 0.0005, the critic learning rate = 0.001,  the coefficient of soft updates  $  \tau = 0.05 $ , the discount factor $ \gamma = 0.05 $, the buffer capacity = 100 000. The noise is complex additive white  Gaussian with mean equal to zero and variance equal to 0.1.


A comparison between NOMA and OMA sum rates versus the number of users is shown in Fig.~\ref{fig:Comparison between NOMA and OMA sum rates}, where the transmit power Pt is 40 dBm. It is realized that NOMA performs better than OMA since it provides higher sum rate for a number of users less than 16. The reason is that in NOMA there is resource sharing among users since NOMA multiplexes users in the power domain, and thus there is no bandwidth division. Therefore the rate and spectral efficiency are higher. However, in OMA there is no resource sharing and thus the bandwidth is divided among users. Further, when the number of users increases above 16, interference between users increases and thus OMA performs better in this case and provides higher sum rate than NOMA. 

 \begin{figure}[t]
    \centering
    \includegraphics[width=0.6\linewidth]{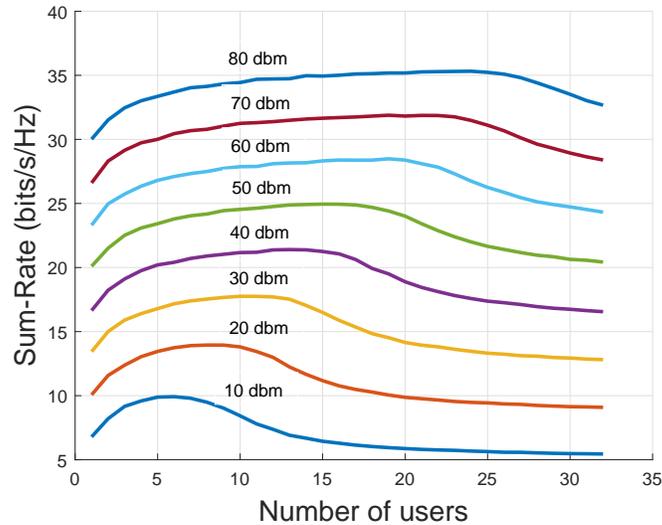}
    \caption{NOMA sum rate for various power levels}
    \label{fig:NOMA sum rate for various power levels}
\end{figure}

 \begin{figure}[t]
    \centering
    \includegraphics[width=0.6\linewidth]{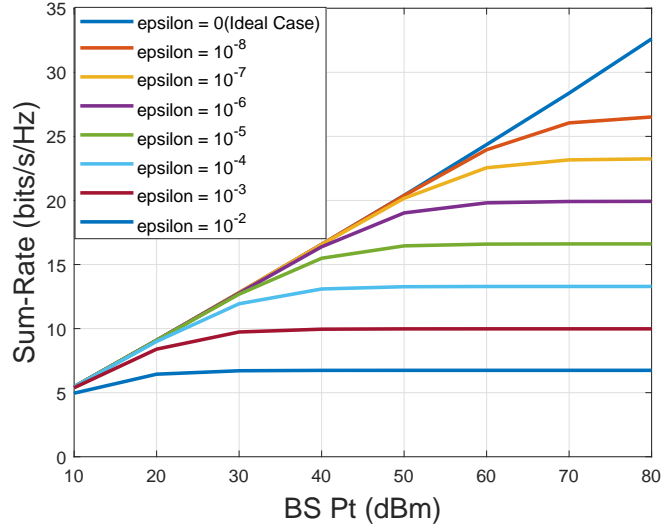}
    \caption{Achievable Sumrate at nearest user with imperfect SIC}
    \label{fig:NOMA sum rate with imperfect SIC}
\end{figure}


Moreover, Fig.~\ref{fig:NOMA sum rate for various power levels}  demonstrates the sum rate of NOMA vs the number of users for different power levels starting from 10 up to 80 dBm. The lower curve represents the sum rate generated at transmit power equals 10 dBm, and the highest curve depicts the sum rate generated at at transmit power equals 80 dBm. It is realized that as power increases the sum rate increases and thus our IRS NOMA system is able to server more number of users.


Furthermore, Fig.~\ref{fig:NOMA sum rate with imperfect SIC} shows the rate for user K, the nearest user to the BS, when considering imperfect SIC. It is well known that user 1, the farthest user from the BS, does not perform SIC and thus we will plot the rate for user K which during imperfect SIC will have residual interference of all users' power in the denominator. It is obvious that as the imperfection increases, the rate will be lower. The curves are plotted for different values of $ \large \epsilon $ which represents the fraction of residual interference. When $ \large \epsilon $ equals 0, SIC is perfect, and thus the rate for user k is the highest. As $ \large \epsilon $ value increases the rate decreases due to increasing the fraction of imperfectness. Therefore, imperfect SIC has a deleterious impact on the rate of the users performing SIC.

\section{Conclusion}

In this paper, we considered the downlink scenario of the IRS NOMA system. Our main goal was to maximize the sum rate of  NOMA users. The formulated problem is non-convex since it involves the constant modulus constraint, and the objective function which is also non-convex. Thus, the problem is suitable for DRL learning techniques. In particular, we have used the DDPG which is a DRL algorithm to solve the sum rate maximization problem for our IRS NOMA scenario. Simulation results revealed that the sum rate for NOMA can track the upperbound obtained through exhaustive search, and it is superior to OMA for a specific number of users and predefined transmit power. Moreover, increasing the transmit power results in increasing the number of users served by the IRS NOMA system since NOMA multiplexes users in the power domain. Further, when considering the imperfect SIC scenario, which is more realistic, results showed that as the imperfection factor increases, the sum rate of users decreases. This reveals the significance of performing SIC perfectly.

\appendices

\ifCLASSOPTIONcaptionsoff
  \newpage
\fi




\end{document}